\documentclass[pre,twocolumn,showpacs,preprintnumbers,floatfix,amsmath,amssymb,superscriptaddress]{revtex4}
\usepackage{graphics}
\usepackage[usenames,dvipsnames]{color}
\usepackage{amsmath,amssymb,bm}
\newcommand{\vf}{v_\text{F}}

\newcommand{\ee}{\text{e}}
\newcommand{\ii}{\text{i}}

\definecolor{NavyBlue}{RGB}{1,101,159}
\definecolor{Orange}{RGB}{255,127,0}

\usepackage[colorlinks=true, citecolor=NavyBlue]{hyperref}

\newcommand{\comment}[1]{}

\begin{document}
\title{Quasiparticle cooling using a Topological insulator-Superconductor hybrid junction}
\author{Dario Bercioux}
\email{dario.bercioux@dipc.org}
\affiliation{Donostia International Physics Center (DIPC), Manuel de Lardizbal 4, E-20018 San Sebasti\'an, Spain}
\affiliation{IKERBASQUE, Basque Foundation of Science, 48011 Bilbao, Basque Country, Spain}

\author{Procolo Lucignano}
\email{procolo.lucignano@spin.cnr.it}
\affiliation{CNR-SPIN, Monte S.Angelo, via Cinthia, I-80126 Napoli, Italy}
\affiliation{Dipartimento di Fisica ``E. Pancini'', Universit\'a di Napoli ``Federico II", Monte S.Angelo, I-80126 Napoli, Italy}
\begin{abstract}
In this work, we investigate the thermoelectric properties of a hybrid junction realised coupling surface states of a three-dimensional topological insulator with a conventional $s$-wave superconductor. 
We focus on the ballistic devices and study the quasiparticle flow, carrying both electric and thermal currents, adopting a scattering matrix approach based on conventional Blonder-Tinkham-Klapwijk formalism. 
We calculate the cooling efficiency of the junction as a function of the microscopic parameters of the normal region (\emph{i.e.} the chemical potential etc.).
The cooling power increases when moving from a regime of  Andreev specular-reflection to a regime where Andreev retro-reflection dominates. Differently  from the case of a conventional  N/S interface, we can achieve efficient cooling of the normal region, without including any explicit impurity scattering at the interface, to increase  normal reflection. 
\end{abstract}
\maketitle
\section{Introduction}
\label{intro}

Semiconductor presenting topological properties as  Bi$_2$Te$_3$, Bi$_2$Se$_3$, CdTe and others have always been considered excellent thermoelectric materials~\cite{Xu:2017} well before the discovery of their topological properties. The most efficient Peltier cells are obtained in PbTe~\cite{Heremans:2008,Biswas:2012} and CdTe~\cite{Kranz:2013}. Recently, there is a great interest in exploring the connection between the  thermoelectric  and topological properties ~\cite{Hasan:2010,Qi:2011,Ando:2013,Bardarson:2013} of  these topological insulator (TI) materials. Two-dimensional or three-dimensional topological insulators (3DTIs) are characterized by conducting one-dimensional   or two-dimensional boundary states, respectively. An aspect of particular interest is the role played by the hybridization of the edge states on different surfaces in influencing the thermoelectric properties~\cite{Murakami:2011,Takahashi:2012}.

In this article, we will explore thermoelectric properties of TIs directly connected to the linear dispersing  surface states of a 3DTI  system. Notably, we will focus on the cooling features of a TI system placed in contact with superconductor (S).  

We investigate an effect based on the property that quasiparticle flow in N/I/S (normal metal/insulator/superconductor junctions), at a bias voltage V, carries not only charge but also heat from N into S. In the latter, due to the presence of a superconducting energy-gap $\Delta$,   quasiparticles with energy exceeding the energy gap ($\epsilon>\Delta$) can tunnel out of N region. This produces a depletion of their energy distribution at high values, thus effectively decreasing the temperature of the N region~\cite{Bardas:1995}.
 Indeed, the heat transfer through N/I/S junctions can be used for the realization of microcoolers \cite{Giazotto:2006}. Present state-of-the-art experiments allow the reduction of the electron temperature in a normal metal lead from 300 to about 100 mK, offering perspectives for on-chip cooling of nano or microsystems \cite{Lowella:2013}. Effects of cooling have been also shown in experimental studies of S/I/N/I/S heterostructure with Al superconductors~\cite{Rajauria:2008}.

Here, we summarize the mechanisms producing this cooling effect~\cite{Bardas:1995}. At low temperatures the charge transport in N/I/S junctions is dominated by a subgap process, the Andreev reflection:  the transfer of a Cooper pair into the superconductor, involving electron to hole conversion at the N/I/S interface.  This process carries no heat current through the interface (because electron-hole energies are symmetric w.r. to the Fermi level). 
By applying a subgap bias across the junctions, the Andreev reflection results in a finite charge current flowing through the N/I/S system. Due to finite resistance of the normal metal, this current generates Joule heating | this is entirely deposited in the normal metal~\cite{Bardas:1995}. Usually, at low temperatures, this heating exceeds the single-particle cooling, and therefore the suppression of Andreev processes is desirable to achieve an efficient cooling.
One way to suppress the Andreev reflection is by decreasing the N/I/S junction transparency. However, large contact resistance hinders hot carrier transfer and leads to a severe limitation in the achievable cooling powers. To increase the junction transparency, and at the same time to reduce the Andreev current, it was suggested to use ferromagnetic metals \cite{Giazotto:2002}, half-metallic ferromagnets \cite{Wolf:2001}, or ferromagnetic insulators \cite{Kawabata:2013}. 

\begin{figure}[!t]
\begin{center}
\resizebox{0.75\columnwidth}{!}{
  \includegraphics{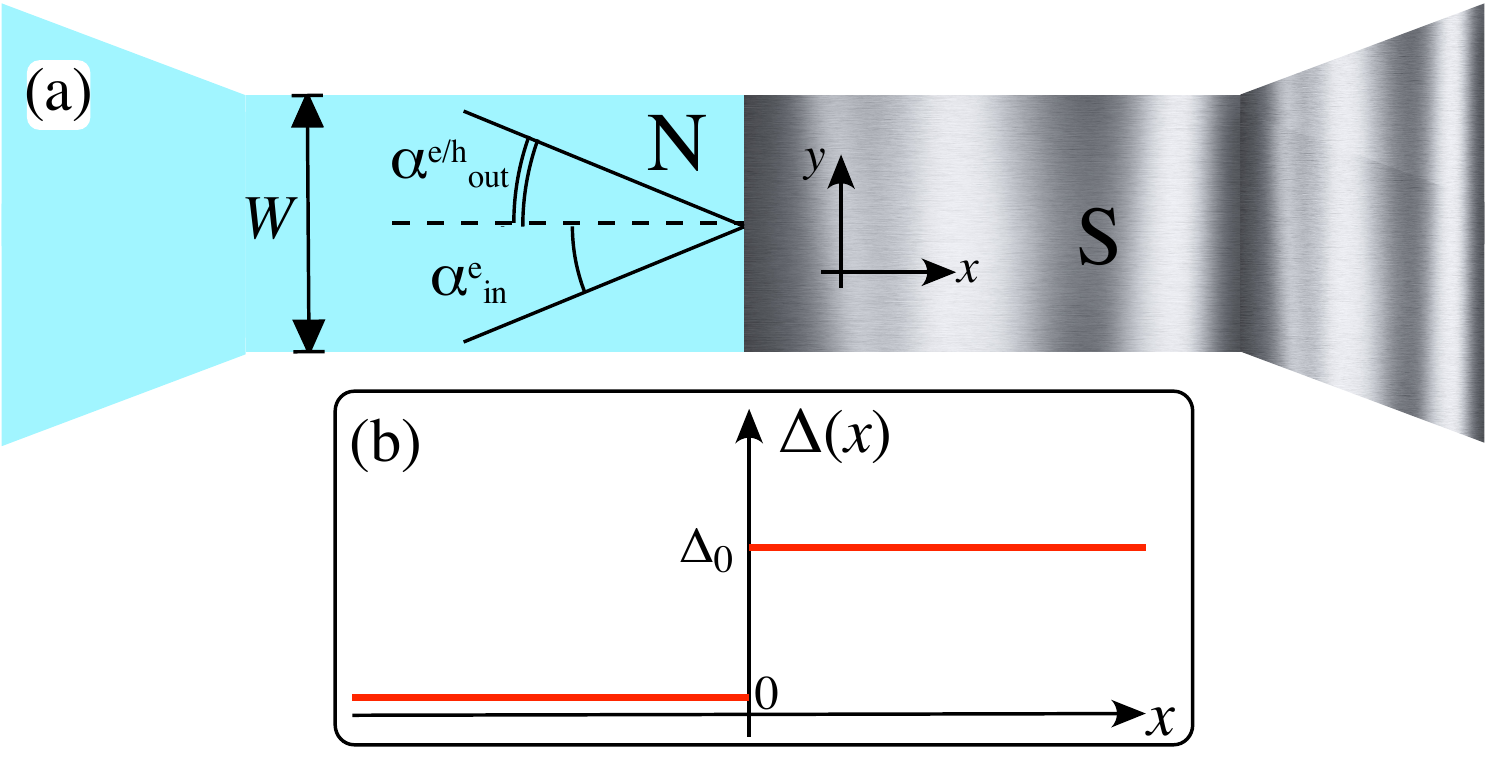} }
\caption{\label{fig:one} (a) Sketch of the N/S interface in the $(x,y)$-plane. (b) The functional behavior of the pairing potential $\Delta(x)$ in Eq.~\eqref{Delta}.}
\end{center}
\end{figure}

In the present paper, we propose a simple N/S junction, where the normal region is the two-dimensional topologically protected surface state of a 3DTI. We will show that we can achieve a finite cooling power for this hybrid junction even in the absence of an insulating region. This effect is possible thanks to the reflection channel that is naturally produced by electrons impinging on the interface with a non-zero incident angle. Furthermore, we show that the cooling power can be modulated by tuning the chemical potential of the normal region. This process corresponds to modify the nature of the Andreev reflection: from retro- to specular-reflection, in a similar fashion to what predicted in single-layer graphene superconducting junctions~\cite{Beenakker:2006}. The set-up we propose [see Fig.~\ref{fig:one}(a)] is similar to the one recently realized experimentally in an N/S junction realized in HgTe 3DTI~\cite{Wiedenmann:2017}.
The effect of a finite barrier was already investigated in Ref.~\cite{Li:2017} in the context of a TI hybrid junction with a mixed singlet and triplet pairing states proximity pairing. In Ref.~\cite{Li:2017} a finite BTK $Z$ parameter induces a cooling power behaving very similarly to the standard BTK case (of Ref.~\cite{Bardas:1995}) with respect to $Z$. In our approach we do not account for a  finite insulating barrier (BTK $Z$ parameter) and show that the Dirac nature of the electrons can give rise to a finite cooling effect as well.

The paper is structured in the following way, in Sec.~\ref{sec:1} we introduce the formalism relative to the normal 3DTI region and the proximitized one. Here we investigate the kinematics of the scattering at the NS interface. We further introduce the expression for the two quantities of interest for this work: the differential conductance and the cooling power. In Sec.~\ref{results} we discuss these quantities for our hybrid junction; specifically, we show that the differential conductance for this system shows strong similarities with those of a  single layer-graphene~\cite{Beenakker:2006}. Then  the cooling power of the hybrid junction is evaluated and discussed. Conclusions and few technical appendices are ending the manuscript.

\section{Model and Formalisms}\label{sec:1}
%
%
%
%
\subsection{The normal region}
We consider an hybrid interface between a normal metal and a superconductor. This is realized on the surface of a 3DTI. We consider the insulator thick enough to neglect the tunneling coupling between the top and bottom surfaces of the system~\cite{Majidi:2016}.
The effective Hamiltonian describing one of the surface edge states of a 3DTI reads:
%
%
\begin{align}\label{gen:ham}
\mathcal{H}_0=  \vf(\bm{\sigma}\times \bm{p})_z\,,
\end{align}
%
%
where $\vf$ is the electron Fermi velocity, $\bm{\sigma}=\{\sigma_x,\sigma_y,\sigma_z\}$ are the Pauli matrices associated to the spin degree-of-freedom, the electron motion is restricted to the $(x,y)$-plane, \emph{i.e.}, $\bm{p}=\{p_x,p_y\}$.  
The Hamiltonian \eqref{gen:ham} describes electrons characterized by a linear energy dispersion:
%
%
\begin{equation}\label{linear}
\mathcal{E}_\pm(\bm{k})=\pm\hbar \vf |\bm{k}|\,.
\end{equation}
%
%
The eigenstates are expressed by the following two-components wave function:
%
%
\begin{align}\label{eigenstates}
\psi_\pm(\bm{k},\bm{r})=\frac{\ee^{\ii \bm{k}\cdot\bm{r}}}{\sqrt{2}}\begin{pmatrix}
1 \\
\mp \ii \ee^{\ii \theta(\bm{k})}
\end{pmatrix}\,,
\end{align}
%
%
where the phase is $\theta(\bm{k})=\arctan(k_y/k_x)$; due to the structure of Hamiltonian~\eqref{gen:ham}, this two-components wave function represents a spin eigenstate. The two states~\eqref{eigenstates} with opposite momenta,  are characterized by opposite spin projections: as well as for a two-dimensional electron gas with Rashba spin-orbit interaction, the two spin eigenstates are always perpendicular to the motion direction $\bm{k}$~\cite{Bercioux:2015}.

%
%
\begin{figure}[!t]
\begin{center}
\resizebox{0.75\columnwidth}{!}{
  \includegraphics{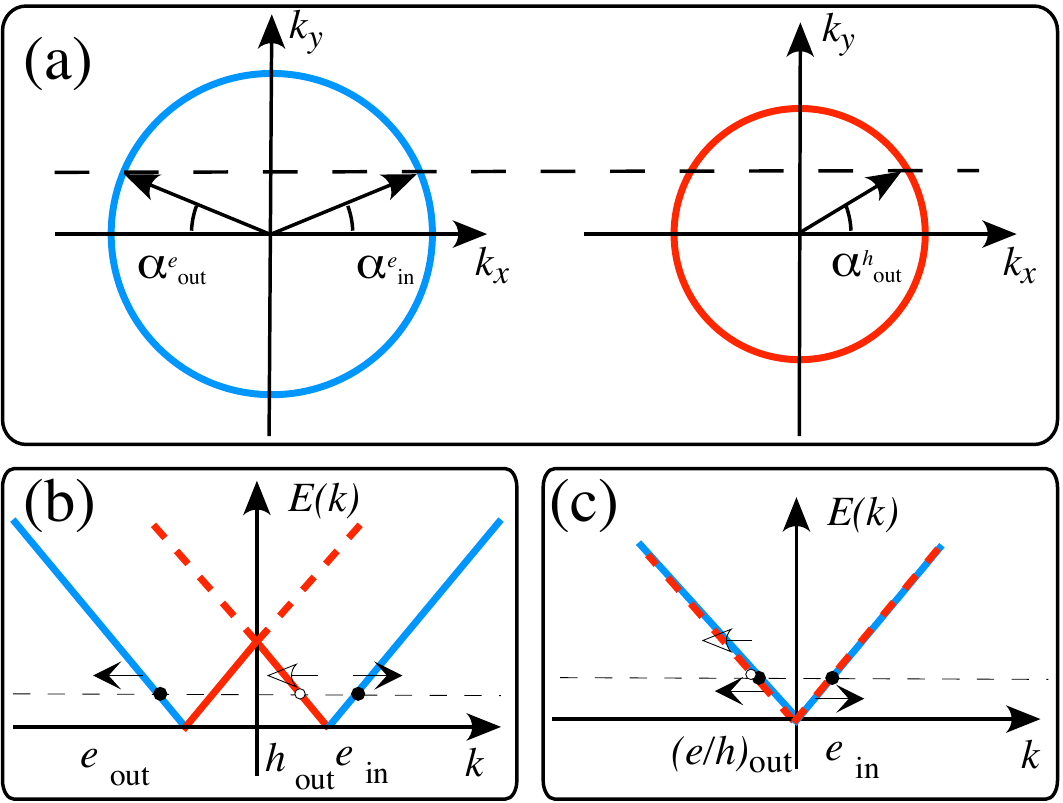} }
\caption{\label{fig:two} Panel (a), Fermi circles for the incoming electrons (left side) and for the Andreev reflected holes (right side). Panel (b) and Panel (c): Sketch of the excitation  spectrum for electron in the conduction band (blue solid line) and holes in the conduction band (red solid line) and valence band (red dashed line) for the case in which $\mu\gg\text{Max}[\Delta,\epsilon]$ and $\mu\ll\text{Max}[\Delta,\epsilon]$, respectively.}
\end{center}
\end{figure}
%
%
\subsection{The superconducting region and the N/S interface}
A superconducting region is created by proximity effect with a $s$-wave superconductor placed on top of the 3DTI~\cite{Fu:2008,Stanescu:2010a,Tkachov:2013b}. Possible 3DTI materials for the set-ups we propose are: HgTe quantum wells~\cite{Wiedenmann:2017}, Bi$_2$Se$_3$~\cite{Galletti:2014} and Bi$_2$Te$_3$~\cite{Veldhorst:2012} multi-layer systems. While non-topological states can have sizable effects on normal conductivity, it has been shown that their contribution to proximity superconductivity seems to be negligible \cite{Veldhorst:2012,Galletti:2014}. Hence, for sake of simplicity,  we ignore them in the following.
In the Nambu space~\cite{deGennesbook}, the full proximitized Hamiltonian of the system reads now:
%
%
\begin{align}\label{ham:Nambu}
\mathcal{H} & = \begin{pmatrix} 
\mathcal{H}_0 -[\mu+U(x)]\sigma_0 & (-\ii \sigma_y) \Delta \ee^{\ii\phi} \\
(\ii \sigma_y) \Delta\ee^{-\ii\phi} & [\mu+U(x)]\sigma_0-\mathcal{T}\mathcal{H}_0 \mathcal{T}^{-1}
\end{pmatrix} 
\end{align}
%
%
where $\sigma_0$ is the identity matrix   in the spin space, $\Delta$ is the superconducting pair potential, $\phi$ the phase of the superconducting phase, $\mu$ is the chemical potential and the time-reversal operator is expressed by $\mathcal{T}=\ii \sigma_y \mathcal{C}$, where $\mathcal{C}$ is the complex conjugation operator. We express  the Hamiltonian of Eq.~\eqref{ham:Nambu} in the following base $\bm{\Psi}=(\Psi_\uparrow,\Psi_\downarrow,\Psi_\uparrow^\dag,\Psi_\downarrow^\dag)$~\cite{Tkachov:2013b,Tkachov:2013a}. In Eq.~\eqref{ham:Nambu}, we used the property that the kinetic term $\mathcal{H}_0$ is time-reversal invariant, \emph{i.e.}, $[\mathcal{H}_0, \mathcal{T}]=0$~\cite{Bercioux:2015}. We have introduced an  electrostatic  potential $U (x)$ in Eq.~\eqref{ham:Nambu} to dope the superconducting region and move away the system from the charge neutrality point~\cite{Beenakker:2006}, in alternative we could consider two different chemical potential $\mu$ for the normal region and the superconducting one~\cite{Cayssol:2008}.

In order to investigate the properties of a N/S interface, in the following we assume a stepwise profile for the superconducting pair potential $\Delta$:
%
%
\begin{align}\label{Delta}
\Delta(x)= \begin{cases}
\Delta_0 & x\ge0 \\
0 & x<0
\end{cases}\,,
\end{align}
%
%
which is not calculated self-consistently throughout the paper.
Our system is therefore characterized by translational invariance along the interface separating the normal$\Delta=0$ from  the superconducting $\Delta\neq0$ region   as sketched  in Fig.~\ref{fig:one}(b). We make  similar  assumptions for the electrostatic potential $U(x)$. Since the zero of potential is arbitrary, we may take
%
%
\begin{align}\label{U}
U(x)=\begin{cases}
U_0 & x\ge0 \\
0 & x<0
\end{cases}.
\end{align}
%
%

This system is very similar to the one considered by Beenakker for the case of single-layer graphene~\cite{Beenakker:2006}, the major differences resides in the consideration in our case of the real spin~\cite{Bercioux:2015} and of the different ordering of the Hamiltonian matrix in the Nambu space.

In Fig.~\ref{fig:one}(a) we present a sketch of the system set-up. The system is translational invariant along the $y$-direction, thus, the $k_y$-component of the momentum is conserved upon electron scattering against the N/S interface. Electrons from the conduction band impinging on the interface with an angle $\alpha^e_\text{in}$ are normal reflected within the same band with opposite angle $\alpha^e_\text{out}=\pi-\alpha^e_\text{in}$ or can be Andreev reflected as hole in the conduction --- retro-reflection --- or in the valence band --- specular-reflection~\cite{Beenakker:2006}. 
Considering an incoming electron in the conduction band, the incidence angle $\alpha_\text{in}$
can be expressed in polar coordinate as:
%
%
\begin{align}\label{injection:angle}
\alpha^{e}_\text{in}& = \arcsin\left[\frac{\hbar \vf k_y}{\epsilon+\mu}\right]\,,
\end{align}
%
%
where $(\epsilon+\mu)(\vf \hbar)^{-1}=|\bm{k}|$ is the modulus of the momentum for the conduction band.
For a hole, the value of the reflection angle can be obtained by simple kinematic considerations: conservation of energy $\epsilon$ and the momentum along the interface $k_y$ gives
%
%
\begin{align}\label{reflection:angle}
\alpha^{h}_\text{out}& = \arcsin\left[\frac{\hbar \vf k_y}{\epsilon-\mu}\right]\,.
\end{align}
%
%
We can treat along the same line of reasoning also the case of a hole as injected state. The propagation direction of the hole strongly depends on the value of the injection energy compared to the chemical potential.
In Fig.~\ref{fig:two}(a) we show a sketch of the the Fermi circles for electrons and holes in the normal region; if the momentum $k_y$ exceeds the radius of the Fermi circle for holes, the corresponding Andreev reflected state cannot propagate. In this case, we introduce a critical injection angle defined as:
%
%
\begin{equation}\label{alphacritical}
\alpha_\text{c} =\arcsin\left[\frac{|\epsilon-\mu|}{\epsilon+\mu}\right]\,. 
\end{equation}
%
%
In order to evaluate the differential conductance and the cooling power of the N/S interface, we need the scattering states in the normal and superconducting regions, a detailed calculation of these scattering states is described in details in App.~\ref{scatteringstates}.

\subsection{The Differential Conductance and the Cooling Power}
In this article we are interested in evaluating two transport properties of the N/S interface related to charge and the heat transport. 
Given a finite transverse dimension $W$, the transverse momentum $k_y$ gets quantized accordingly to $k_y(n) = \frac{\pi(n+1/2)}{W}$ where $n\in\mathbb{N}$~\cite{Bardarson:2013}.
The zero temperature differential conductance~\cite{Beenakker:2006,Blonder:1982,Mortensen:1999} is defined as:
%
%
\begin{align}\label{dIdV0}
\frac{\partial I}{\partial V} = \frac{2e^2}{h}\sum_{n=1}^{N} \left( 1+ |r_\text{A}(eV,\alpha_n)|^2-|r_\text{N}(eV,\alpha_n)|^2 \right)\,,
\end{align}
%
%
where 
%
%
\begin{align}\label{reflection:angle}
\alpha_n& = \arcsin\left[\frac{\hbar \vf k_y(n)}{\epsilon+\mu}\right]\,.
\end{align}
%
%
and $r_\text{A}(eV,\alpha_n)$ and $r_\text{N}(eV,\alpha_n)$ are Andreev and normal reflection amplitudes, respectively (see App.~\ref{scatteringproblem} for their explicit calculation).
In the limit  of wide junctions the spacing between different transverse modes can be considered negligible, and we can recast the sum into an integral over the (almost continuous) angle $\alpha_n$ using the following transformations:
%
%
\begin{widetext}
\begin{align}\label{modetoangle}
\sum_n \to \frac{W}{2\pi}\int_{-\infty}^\infty d k_y = \frac{W}{2\pi}\int_{-\pi/2}^{\pi/2} \frac{\epsilon+\mu}{\hbar \vf}\cos\alpha d\alpha=\frac{1}{2}\int_{-\pi/2}^{\pi/2} \mathcal{N}(\epsilon)\cos\alpha d\alpha\,.
\end{align}
Hence:
%
%
\begin{align}\label{dIdV}
\frac{\partial I}{\partial V} = G_0(eV)\int_0^{\pi/2} &\!\!\!d\alpha \cos\alpha\left( 1+ |r_\text{A}(eV,\alpha)|^2-|r_\text{N}(eV,\alpha)|^2 \right)\,,
\end{align}
%
%
where $G_0(eV)=\frac{2e^2}{h}\mathcal{N}(eV)$ represents ballistic conductance of the surface states, and   $\mathcal{N}(\epsilon)=\frac{W}{\pi}\frac{\mu+\epsilon}{\hbar \vf}$ is the number of active transverse modes at the energy $\epsilon$.
Working along the same route we can define the electronic heat current~\cite{Bardas:1995,Kawabata:2013,Leivo:1998,Rouco:2018} at the temperature $T$ and at the voltage $V$ as:
%
%
\begin{align}\label{coolingpower}
\mathcal{J}(T,V)=& \frac{1}{\pi \hbar} \int_{-\infty}^\infty d\epsilon \int_0^{\pi/2} d\alpha \cos\alpha \,\,\mathcal{N}(\epsilon)\left\{ \left[\epsilon(1-|r_\text{A}(\epsilon,\alpha)|^2-|r_\text{N}(\epsilon,\alpha)|^2)\right]\right. - \nonumber \\
&\left. eV\left[\left(1+|r_\text{A}(\epsilon,\alpha)|^2-|r_\text{N}(\epsilon,\alpha)|^2\right)\right]\right\} \left[f_\text{N}(\epsilon-eV,T)-f_\text{S}(\epsilon,T)\right]\,,
\end{align}
\end{widetext}
%
%
where $f_\text{N(S)}$ are the Fermi functions describing the electrons in the normal (superconducting) region.
In the result section we are going to express the cooling power in units of $\Delta^2/e^2R_\text{N}$, where $R_\text{N}$ is $contact~ resistance$, defined as the inverse of $G_0(0)=\frac{2e^2}{h}\frac{W}{\pi}\frac{\mu}{\hbar \vf}=\frac{4e^2}{h^2}\frac{W\mu}{\vf}$ that is the quantum of conductance times the number of open channels at the Fermi levels.

The general expression for the scattering amplitudes calculated in App.~\ref{scatteringproblem} are given by: 
%
%
\begin{widetext}
\begin{subequations}\label{results0}
\begin{align}
r_\text{A} &= \begin{cases} 
\ee^{\frac{\ii}{2}(\alpha+\alpha'+2\phi-\pi)}X^{-1}\sqrt{\cos(\alpha)\cos(\alpha')} & \text{if}~|\alpha|<\alpha_\text{c}\\
0 & \text{if}~|\alpha| > \alpha_\text{c}
\end{cases} \, \\
r_\text{N} & = \ee^{\frac{\ii}{2} (2\alpha-\pi)} X^{-1} \left[\cos\beta \sin\left(\frac{\alpha+\alpha'}{2}\right) + \ii \sin\beta\sin\left(\frac{\alpha-\alpha'}{2}\right)\right]\,, \\
r_\text{A}'& = -\ee^{2\ii \phi} r_\text{A}\,,\\
r_\text{N}'&= \ee^{\frac{\ii}{2} (\alpha'-\pi)} X^{-1} \left[\cos\beta\sin\left(\frac{\alpha+\alpha'}{2}\right)-\ii \sin\beta\sin\left(\frac{\alpha-\alpha'}{2}\right)\right]\,, \\
X& = \cos\left(\frac{\alpha+\alpha'}{2}\right)\cos\beta+\ii \cos\left(\frac{\alpha+\alpha'}{2}\right)\sin\beta\,.
\end{align}
\end{subequations}
\end{widetext}
%
%
In the previous expression we have simplified  the notation from $\alpha_\text{in}^e\to\alpha$ and $\alpha_\text{out}^h\to\alpha'$.
Here the ``prime" index on $r_\text{N}$ and $r_\text{A}$ indicates the \emph{inverse} reflection upon change of the injected charge, \emph{e.g.}, an injected hole instead of an electron. 
The previous quantities are obtained by assuming that the potential $U_0$ is large enough to dope the superconducting region far away from the charge neutrality point of the Dirac dispersion in Eq.~\eqref{U}. In this regime, we can safely work in the \emph{Andreev approximation}~\cite{deGennesbook,Beenakker:2006,Cayssol:2008,Linder:2008}.
In the Eqs.~\eqref{results0} we have introduced the standard superconducting function $\beta(\epsilon)$ defined as
%
%
\begin{align}\label{beta}
\beta(\epsilon)=\begin{cases}
\arccos\left(\frac{\epsilon}{\Delta}\right) & \epsilon<\Delta \\
-\ii\, \text{arccosh}\left(\frac{\epsilon}{\Delta}\right) & \epsilon>\Delta
\end{cases}\,.
\end{align}
%
%
With the amplitudes defined in Eqs.~\eqref{results0}, we can introduce an interface scattering matrix:
%
%
\begin{equation}\label{sm:NS}
\mathcal{S}_\text{NS}=\begin{pmatrix}
r_\text{N} & r_\text{A}' \\
r_\text{A}& r_\text{N}'
\end{pmatrix}\,,
\end{equation}
%
%
that is unitary $\mathcal{S}^\dag_\text{NS} \mathcal{S}_\text{NS}=\mathcal{S}_\text{NS} \mathcal{S}^\dag_\text{NS}=\mathbb{I}_2$ for energies smaller than the gap $\Delta$. It is worth noticing that the expressions~\eqref{results0} are similar to the analogous ones obtained for the case of single-layer graphene in Ref.~[\onlinecite{Beenakker:2006}].

We consider now two limiting cases for the reflections amplitudes \eqref{results0}: If the chemical potential is the dominating energy scale $\mu\gg\text{Max}[\Delta,\epsilon]$, then the hole generated during the Andreev refection results in an empty states in the conduction band that has a propagation angle that is opposite compared to the incoming electron $\alpha_\text{out}^h=-\alpha^e_\text{in}$ [$\alpha'=-\alpha$ in Eqs.~\eqref{results}]. This is the limit of Andreev retro-reflection, here we can write a simplified expression for the Andreev and the normal reflection amplitudes [c.f. Fig.~\ref{fig:two}(b)]:
%
%
\begin{subequations}\label{retro}
\begin{align}
r_\text{A}& =\frac{\ii\ee^{\ii\phi}\cos\alpha}{\zeta + \frac{\epsilon}{\Delta}\cos\alpha}\,,\label{AR} \\
r_\text{N}& = \frac{\ii\ee^{\ii \alpha}\zeta\sin\alpha}{\zeta + \frac{\epsilon}{\Delta}\cos\alpha}\,. \label{NR}
\end{align}
\end{subequations}
%
%
In the opposite regime, $\mu\ll\text{Max}[\Delta,\epsilon]$, we have that the hole generated in the Andreev reflection is in the valence band and that $\alpha_\text{out}^h=\alpha^e_\text{in}
$  [$\alpha'=\alpha$ in Eqs.~\eqref{results0}] --- this is the case of Andreev specular-reflection.~\cite{Beenakker:2006} As for the previous case we can strongly simplify the Eqs.~\eqref{results0} for the normal and Andreev reflection amplitudes [c.f. Fig.~\ref{fig:two}(c)]:
%
%
\begin{subequations}\label{specular}
\begin{align}
r_\text{A}& =\frac{\ii\ee^{\ii(\alpha-\phi)}\cos\alpha}{\frac{\epsilon}{\Delta} +\zeta \cos\alpha}\,,\label{AS} \\
r_\text{N}& = \frac{\ii\ee^{\ii \alpha}\frac{\epsilon}{\Delta}\sin\alpha}{\frac{\epsilon}{\Delta} +\zeta \cos\alpha}\,. \label{NS}
\end{align}
\end{subequations}
%
%
In the two limiting cases presented in Eqs.~\eqref{retro} and \eqref{specular}, we have introduced the following  function:
%
%
\begin{align}\label{zeta}
\zeta=\begin{cases}
\ii \sqrt{1-\left(\frac{\epsilon}{\Delta}\right)^2} & \epsilon < \Delta \\
\sqrt{\left(\frac{\epsilon}{\Delta}\right)^2-1} & \epsilon>\Delta
\end{cases}\,.
\end{align}
%
%
We note that the previous limiting results are compatibles with the ones by Beenakker in Ref.~[\onlinecite{Beenakker:2006}] for the case of single-layer graphene.

\section{Results}\label{results}
\begin{figure}[!t]
\begin{center}
\resizebox{\columnwidth}{!}{
  \includegraphics{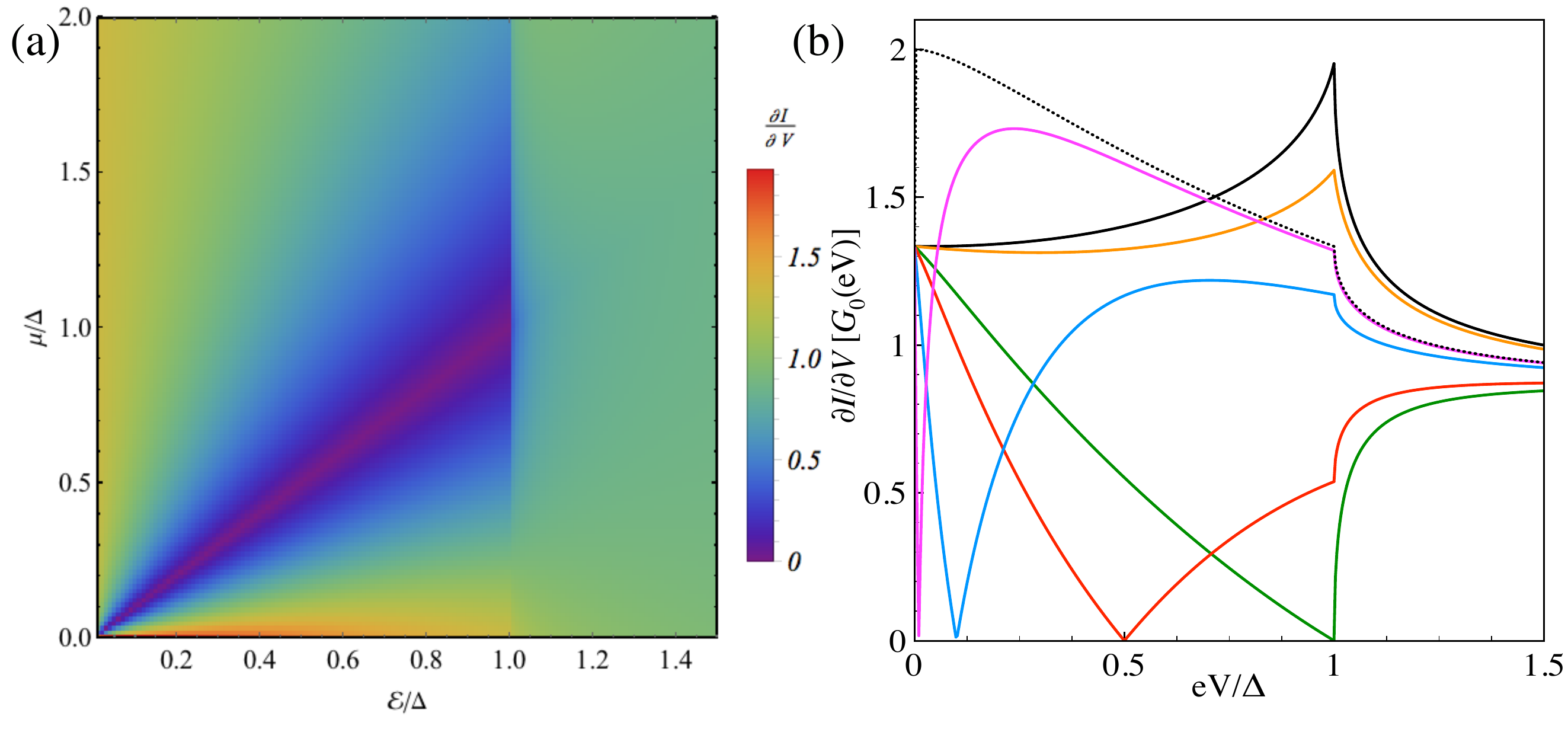} }
\caption{\label{fig:conducatnce} (a) Density plot of the differential conductacnce as a function of the chemical potential $\mu$ and the injection energy $\epsilon$. (b) Differential conductance as a function of the applied voltage at the N/S interface for different values of the chemical potential $\mu$: $\mu=100\Delta$ (solid black line), $\mu=10\Delta$ (solid orange line), $\mu=1\Delta$ (green solid line), $\mu=0.5\Delta$ (red solid line), $\mu=0.1\Delta$ (blue solid line), $\mu=0.01\Delta$ (magenta solid line) and $\mu=0$ (black dotted line). The two limiting cases of Andreev retro- and specular-reflection are presented with black lines: solid and dotted, respectively.}
\end{center}
\end{figure}

\begin{figure}[!t]
\begin{center}
\resizebox{\columnwidth}{!}{
  \includegraphics{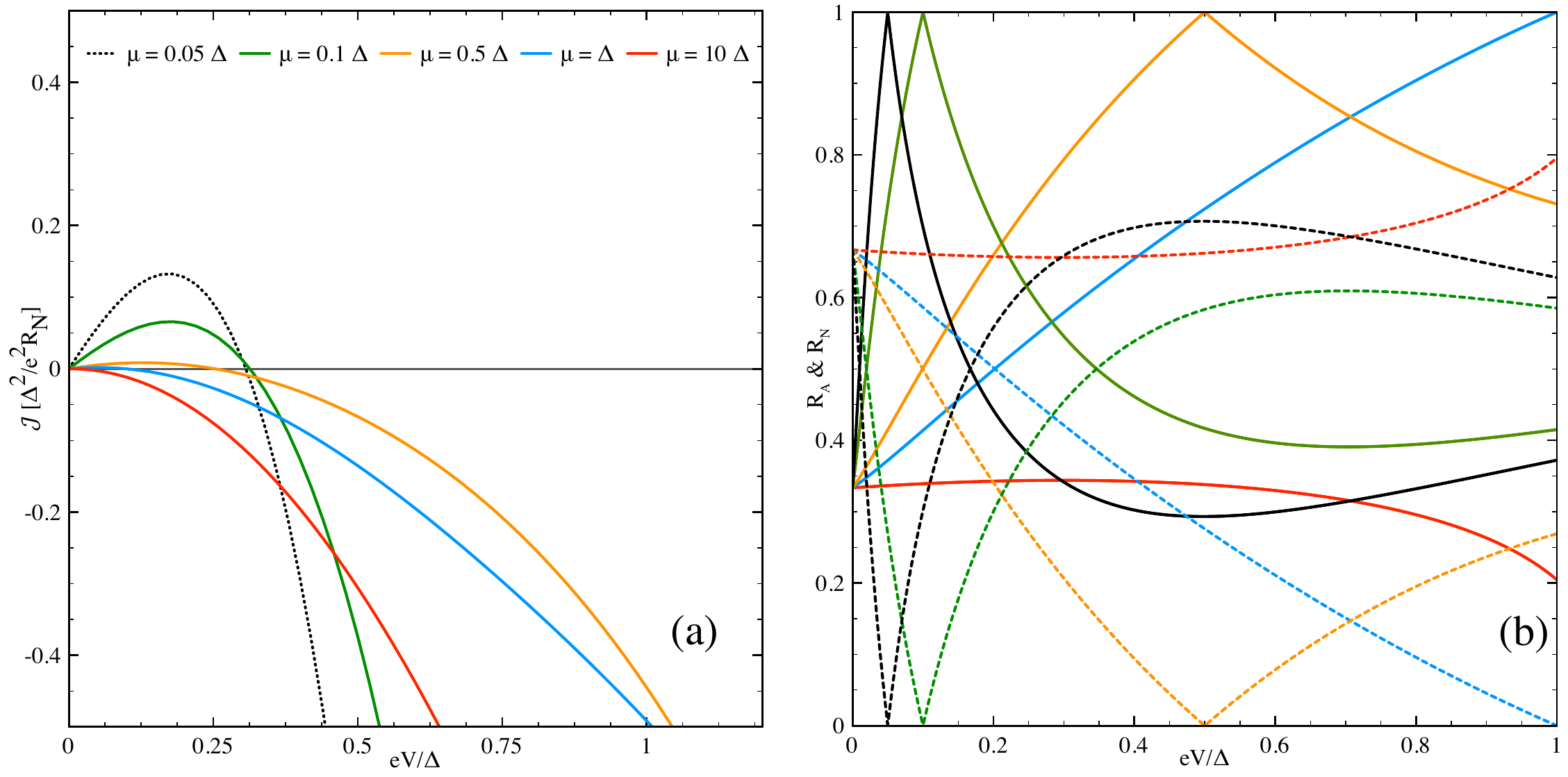} }
\caption{\label{fig:four} Panel (a): Cooling power as a function of the applied voltage for different values of the chemical potential $\mu$ at $T=0.5T_\text{c}$. Panel (b): Andreev reflection (dashed lines) and Normal reflection (solid lines) as a function of the applied voltage for the same values of the chemical potential as in panel (a).}
\end{center}
\end{figure}

We first  present the results for the differential conductance \eqref{dIdV}. In Fig.~\ref{fig:conducatnce} we show the zero temperature differential conductance parametrised as a function of the chemical potential. We show the two limiting cases of Andreev retro-reflection $\mu\gg\max[\Delta,\epsilon]$, c.f.~Eq.~\eqref{retro}, and the opposite case of Andreev specular-reflection $\mu\ll\max[\Delta,\epsilon]$ --- c.f. Eq.~\eqref{specular}. Our results are in agreement with those obtained for the case of single-layer graphene~\cite{Beenakker:2006}, but do not coincide with the ones of Ref.~\cite{Majidi:2016} for a 3DTI.
%
%
%
%
In the limit of Andreev retro-reflection, the differential conductance at zero applied bias is equal to $4/3G_0(eV)$, than increases to a value of $2G_0(eV)$ at the gap $eV=\Delta$, whereas in the limit of Andreev specular-reflection is equal to $2G_0(eV)$ at zero applied bias and it decreases to $4/3G_0(eV)$ for an applied voltage equal to the gap. In the limit of very large applied voltages,  $eV\gg\Delta$, both in the retro and in the specular limit, the conductance has a limiting value of $(4-\pi)G_0(eV)$. These values can be easily obtained by using the expressions \eqref{retro} and \eqref{specular} and the definition of differential conductance~\eqref{dIdV}.  These behaviors of the differential conductance differ form the ones known in the case of  standard NIS interfaces~\cite{Blonder:1982}.  Here the Dirac nature of the injected electrons in the 3DTI normal region is a natural source of normal reflection for injection angle different from zero. The integrated normal and Andreev reflections for sub-gap injection energies are shown in Fig.~\ref{fig:four}(b). We observe that for the intermediate values of the chemical potential $\mu\le\Delta$ the differential conductance is zero for $\epsilon=\mu$ [c.f. Fig.~\ref{fig:conducatnce}(a)], the reason is that for this energy the Andreev reflection is zero  --- because the density of states is zero for the holes [c.f. Fig.~\ref{fig:two}(b)] --- and, consequently, the normal reflection is equal to one. The zero represents a crossover from a regime of almost Andreev retro-reflection ($\epsilon<\mu$) to a regime of almost Andreev specular-reflection ($\epsilon>\mu$). It is worth to note that complete Andreev retro- and specular-reflections are recovered only in the limiting cases of $\mu\gg\text{Max}[\Delta,\epsilon]$ and $\mu\ll\text{Max}[\Delta,\epsilon]$, respectively. 
%
%
%
%

\subsection{Cooling power}

In this section we present results for the cooling power defined in Eq.~\eqref{coolingpower}.
We are going to show that it strongly depends on the type of Andreev reflection. 
In discussing our results, we mostly focus on the case $T_\text{bath} = 0.5 T_\text{c}$.  At these temperatures, deviations of the superconducting gap $\Delta(T)$, w.r. to the zero temperature gap $\Delta(0)$ should be taken into account solving the self-consistent gap equation at each temperature. In our calculation, we do not perform this calculation and assume the gap to be equal to the zero temperature  gap. Of course, this limits the accuracy of our results at high temperatures --- close to $T_\text{c}$.
 In the case of Al superconducting electrodes this corresponds to a temperature of $\sim 500$~mK.   
Hence, our predictions could be easily measured in conventional He$_3$ refrigerators.
The behavior of the cooling power as a function of the applied voltage for different values of the ratios $\mu/\Delta$ is shown in Fig.~\ref{fig:four}(a). 

Surprisingly, we find that it can be sizable, at low voltages, in the regime of Andreev specular-reflection (\emph{i.e.} when the chemical potential $\mu\ll\text{Max}[\Delta,\epsilon]$) while it is always negligible in the Andreev retro-reflection  regime ($\mu\gg\text{Max}[\Delta,\epsilon]$). This result is remarkable, because tuning the chemical potential via an external gate, we can explore both regimes.
The curves as a function of the applied voltage recall  those obtained for a conventional normal region obtained in Ref.~\cite{Bardas:1995}. 
Our main  difference  w.r. to Ref.~\cite{Bardas:1995,Kawabata:2013,Rouco:2018} is that we are considering a clean interface between the normal and the superconducting region. 
By contrast in a conventional metal, with Schr\"odinger electrons with  parabolic dispersion, a finite cooling power is achieved if and only if the N/S interface is not clean, \emph{i.e.} it presents a finite $Z$ parameters in the Blonder-Tinkham-Klapwijk derivation~\cite{Bardas:1995,Blonder:1982}. The role of an insulating interface is reported in Ref.~\cite{Li:2017} for a topological system.
%
%
\begin{figure}[!b]
\begin{center}
\resizebox{\columnwidth}{!}{
  \includegraphics{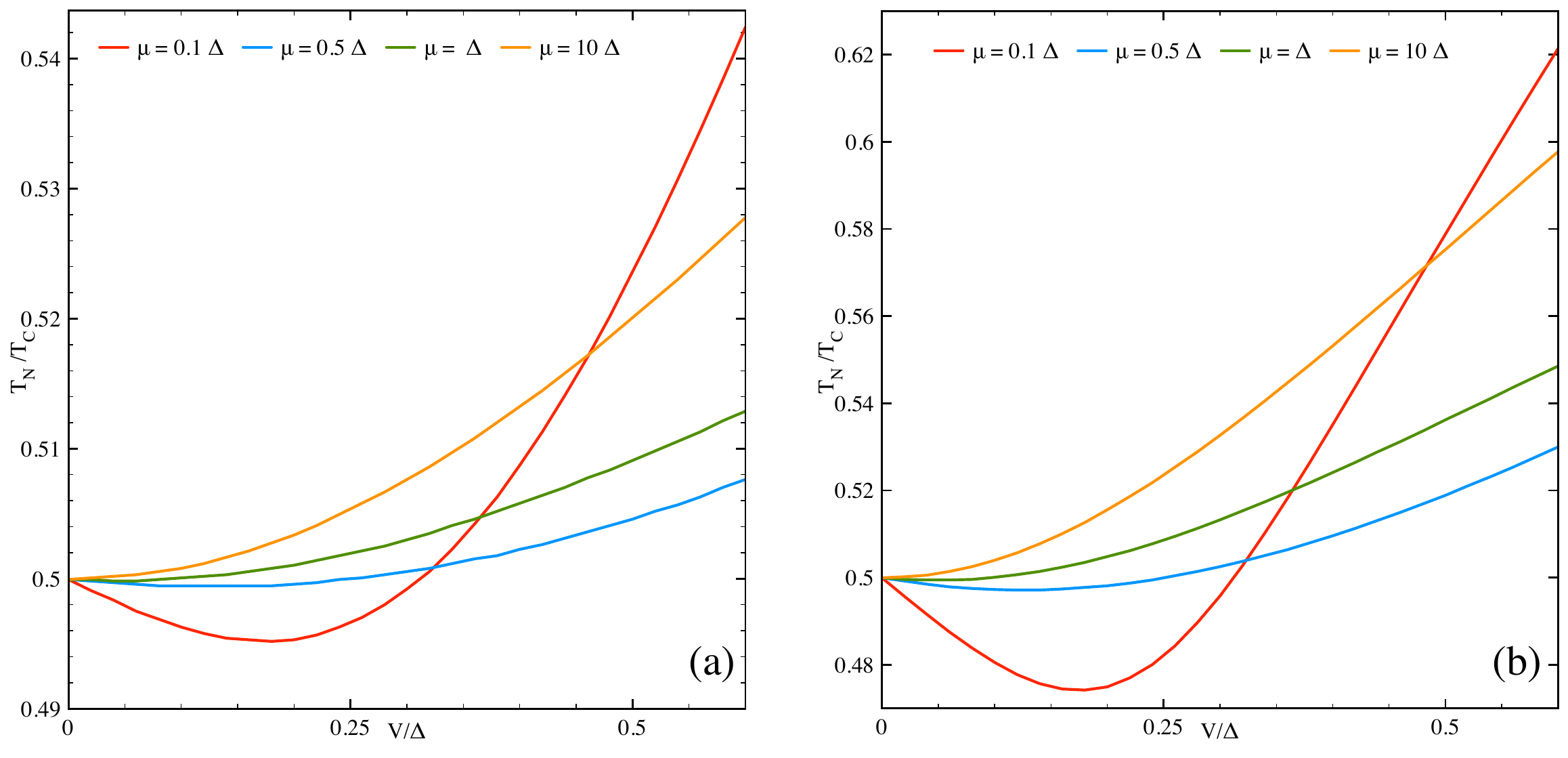}}
\caption{\label{figure5} Amount of cooling of the normal region as a function of the applied voltage and for various values of the chemical potential for two different coupling strengths to the phonon bath: Panel (a) $\mathcal{N}(0)=20$ and Panel (b) $\mathcal{N}(0)=100$. }
\end{center}
\end{figure}
%
%
To  further investigate on these results, in Fig.~\ref{fig:four}(b) we plot the Andreev and normal reflection probabilities as a function of the voltage for the same values of the ratios $\mu/\Delta$ plotted in Fig.~\ref{fig:four}(a).  Interestingly enough, the cooling power is positive and sizable when the Andreev reflection probability (dashed line) is suppressed and the corresponding  normal reflection (solid line) is moderate, though finite and smaller than the former, over almost all the voltage range.  This occur at very small ratios $\mu/\Delta$ corresponding to black and green curves in Figs.~\ref{fig:four}(a) and \ref{fig:four}(b) . By further increasing this ratio (moving versus the retro-reflection regime) the Andreev process becomes dominant, and the cooling effect vanishes. Hence, we expect a sizable cooling effect when the chemical potential of the surface state is finely tuned very close to the charge neutrality point.
We expect to recover cooling power in the retro-reflection regime considering the presence of an insulating interface between the normal region and the superconducting one~\cite{Li:2017}.

For better characterizing the cooling effects on the device addressed, we also study  the temperature decrease of the N-region, adopting a simple energy transfer  balancing model. 
In particular, the heat transported by the electrons in the metal is transmitted to the bath in  the superconducting region which is assumed to be in  equilibrium with the  thermal bath via the phonons. Hence $T_\text{S}=T_\text{ph}=T_\text{bath}$, where $T_\text{S}$ is the superconductor temperature, $T_\text{ph}$ is the phononic temperature and $T_\text{bath}$ is the bath temperature.  The effective electron temperature is determined by the rate at which the electrons can transfer energy to the phonons. It is given by $\dot Q_{e-\text{ph}}=\Sigma \mathcal{V}(T_e^5-T_\text{ph}^5)$, where $\mathcal{V}$ is the volume of the normal region (assumed to be $\mathcal{V}\sim 0.5~\mu$m$^2$ in the following),   $T_{e}$ is the effective electronic temperature, and $\Sigma$ is a material dependent parameter (for Al $\Sigma\sim 0.2-0.3 \cdot 10^9$ W m$^{-3}$ K$^{-5}$). The electronic temperature $T_e$ is the solution of the equation:
%
%
\begin{align}\label{cooling_temperature1}
\mathcal{J}(T_e,T_\text{ph},V)+\Sigma \mathcal{V}(T_e^5-T_\text{ph}^5)=0
\end{align}
%
%
where 
%
%
\begin{widetext}
\begin{align}\label{cooling_temperature2}
\mathcal{J}(T_e,T_\text{ph},V)=& \frac{1}{\pi \hbar} \int_{-\infty}^\infty d\epsilon \int_0^{\pi/2} d\alpha \cos\alpha \,\,\mathcal{N}(\epsilon)\left\{ \left[\epsilon(1-|r_\text{A}(\epsilon,\alpha)|^2-|r_\text{N}(\epsilon,\alpha)|^2)\right]\right. - \nonumber \\
&\left. eV\left[\left(1+|r_\text{A}(\epsilon,\alpha)|^2-|r_\text{N}(\epsilon,\alpha)|^2\right)\right]\right\} \left[f_\text{N}(\epsilon-eV,T_e)-f_\text{S}(\epsilon,T_\text{ph})\right]\,.
\end{align}
\end{widetext}
%
%
This expression is the same  as in Eq.~\eqref{coolingpower}, however here we consider different temperature for the normal and the superconducting regions. 

Equation~\eqref{cooling_temperature1} depends not only on the way electrons couples to the bath via the phonons but also on the number $\mathcal{N}$ of conducting channels in the normal regions.

In Figs.~\ref{figure5}(a) and \ref{figure5}(b), we show the effective electron temperature at fixed  $\Sigma \mathcal{V}$ in the case of $\mathcal {N}(0)=20$ (panel a) and $\mathcal{N}(0)=100$ (b) open channels. Of course, increasing the number of open channels improves the cooling efficiency, as  the relative weight of electron cooling w.r. to phonon heating in Eq.~\eqref{cooling_temperature1} increases. 
Inspection of Fig.~\ref{figure5} shows that our  hybrid interface can effectively cool down the normal region of between above the  20\%  of the initial temperature in the cases of small chemical potential in the more efficient case of $\mathcal{N}(0)=100$. A more detailed explanation of the evaluation of the  effective electronic temperature in reported in App.~\ref{rate:eq}.

\section{Discussion and Conclusions}

To summarize, in this paper we show that an NS interface, in which the normal region is given by the surface states of a 3D topological insulator, can be efficiently used as a quasiparticle cooler even in the absence of an insulating interface~\cite{Bardas:1995}. 
Conventional NIS microcoolers are tuned by varying the transparency of the insulating barrier between N and S. In our case, we can achieve the same tunability, by changing the value of the chemical potential in the normal region with respect to the charge neutrality point of the 3DTI. Recent proposals in this direction were already put forward  to achieve efficient thermal switches | these are based on topological Josephson junctions~\cite{Sothmann:2016,Sothmann:2017}. However, the role played by the chemical potential was not at the center of the investigation like in the present research work. As a possible follow-up, we are planning to understand how our results are modified in cylindrical wire configuration with and without the presence of a magnetic field parallel to the axes of the tube, similarly to the geometry introduced in~\cite{Bardarson:2013}. Furthermore, following  Refs.~\cite{Schwab:2011,Culcer:2012}, we could include the  effects of inhomogeneities and disorder into our modeling.

\section{Acknowledgments}
Discussions with S. Bergeret, B. Bujnowski, P. Burset, A. Tagliacozzo and B. Trauzettel are acknowledged. 
Our work is supported by Spanish Ministerio de Econom\'ia y Competitividad (MINECO) through the projects  FIS2014-55987-P and FIS2017-82804-P. DB is also supported by the Transnational Common Laboratory \emph{QuantumChemPhys}. 
\appendix

\section{The scattering states}\label{scatteringstates}

\subsection{The normal region}\label{snormal}
The scattering states of the normal region are obtained by finding the eigenstates of the Hamiltonian \eqref{ham:Nambu} for $\Delta=0$, we express them as:
%
%
\begin{subequations}\label{scatN}
\begin{align}
\Psi_e^\rightarrow & = \frac{1}{\sqrt{2|\cos(\alpha_e)|}}\ee^{\ii (k_yy+k^e_x x)}(1,-\ii \ee^{\ii \alpha_e},0,0)^\text{T} \\
\Psi_e^\leftarrow & = \frac{1}{\sqrt{2|\cos(\alpha_e)|}}\ee^{\ii (k_yy-k^e_xx)}(1,\ii \ee^{-\ii \alpha_e},0,0)^\text{T}
\end{align}
\begin{align}
\Psi_h^\rightarrow & =\frac{1}{\sqrt{2|\cos(\alpha_h)|}} \ee^{\ii (k_yy+k^h_xx)}(0,0,1,\ii \ee^{\ii \alpha_h})^\text{T} \\
\Psi_h^\leftarrow & =\frac{1}{\sqrt{2|\cos(\alpha_h)|}} \ee^{\ii (k_yy-k^h_xx)}(0,0,1,-\ii \ee^{-\ii \alpha_h})^\text{T} 
\end{align}
\end{subequations}
%
%
where the coefficient $\alpha$ and $k$ contained the scattering states are defined as:
%
%
\begin{subequations}\label{defN}
\begin{align}
\alpha_{e/h}& = \arcsin\left[\frac{\hbar \vf k_y}{\epsilon\pm\mu}\right] \\
k_{e/h} & = \frac{\epsilon\pm\mu}{\hbar \vf}\cos(\alpha) \label{k:vec}
\end{align}
\end{subequations}
%
%
Energy eigenvalues for electrons and holes are 
%
%
\begin{subequations}\label{spectra}
\begin{align}
\mathcal{E}_{e/h}  = (\pm\hbar\vf \sqrt{k_x^2+k_y^2}\mp\mu)  
\end{align}
\end{subequations}
%
%
where for both expressions the $\pm$ sign refers to helicity. We define here the helicity~\cite{Weinberg:1995} as 
%
%
\begin{align}
\hat{h}= \left(\bm{\sigma}\times\frac{\bm{p}}{|\bm{p}|}\right)\,,
\end{align}
%
%
 this operator tell us about the how the spin is rotating along the Fermi surface. Thus, we note from  Eqs.~\eqref{scatN} that the change of direction motion along the $x$ axes ($k\to-k$) corresponds also to a change of helicity.
\subsection{The superconducting region}\label{ssuper}

In the superconducting region we look for solution of the equations 
%
%
\begin{equation}\label{sol:ham}
[\mathcal{H}-\epsilon\mathbb{I}_4]\begin{pmatrix}\bm{u}\\\bm{v}\end{pmatrix}\ee^{\ii(k_yy\pm kx)}=0\,.
\end{equation}
%
%
First we will look for a solution in the simplified regime of a very high doping $U$ for the superconducting region. In this limit, the propagation angle inside the superconducting region
%
%
\begin{align}\label{gamma}
\gamma & = \arcsin\left[\frac{\hbar \vf k_y}{\epsilon+\mu+U}\right] 
\end{align}
%
%
can be neglected $\gamma\to0$ that is equivalent to quasi-electron and quasi-hole states propagating both along the same direction with angle equal to zero --- this is equivalent to the Andreev approximation performed only in the superconducting region. 
By solving Eq.~\eqref{sol:ham}, this system of equations and shifting in an opportune way the coefficients $\bm{u}$ and $\bm{v}$, we find the following solutions for left- and right-moving quasi-electrons:
%
%
\begin{equation}\label{qe}
\psi_\text{QE}^\rightleftharpoons= \ee^{\ii (k_yy\pm k_x^\text{QE}x)}\left(\ee^{\ii \beta}, \mp\ii \ee^{\ii\beta}, \ii \ee^{-\ii\phi} \ee^{\pm \ii\gamma},\pm \ee^{-\ii \phi} \right)^\text{T}
\end{equation}
%
%
whereas for left- and right-moving quasi-holes we find:
%
%
\begin{equation}\label{qh}
\psi_\text{QH}^\rightleftharpoons= \ee^{\ii (k_yy\mp k_x^\text{QH} x)}\left(\ee^{-\ii \beta}, \pm\ii \ee^{-\ii\beta},\ii \ee^{-\ii\phi}  ,\mp \ee^{-\ii \phi} \right)^\text{T}
\end{equation}
%
%
where the function $\beta(\epsilon)$ was defined in Eq.~\eqref{beta} and the longitudinal momentum is defined as:
%
%
\begin{subequations}\label{defS}
\begin{align}
k_x^\text{QE/QH} & = \frac{1}{\hbar \vf}\sqrt{\left[\sqrt{\epsilon^2-\Delta^2}-(U+\mu)\right]^2-k_y^2}.
\end{align}
\end{subequations}
%
%
We can clearly see that the momentum $k$ can have an imaginary part if $\epsilon<\Delta$, while in the opposite case in real.  

\section{The scattering problem}\label{scatteringproblem}

We consider the following two processes: 
%
%
\begin{enumerate}
\item one right-moving electron, reflected into a left-moving electron [normal reflection] and a left-moving hole [Andreev reflection] and transmitted into the superconductor as a linear combination of a right-moving quasi-electron and right-moving quasi hole:
%
%
\begin{equation}\label{P1}
\left.\psi_e^\rightarrow+r_\text{N}\psi_e^\leftarrow+r_\text{A}\psi_h^\rightarrow\right|_{x=0}=\left.a\psi_\text{QE}^\rightarrow+b\psi_\text{QH}^\rightarrow\right|_{x=0}\,;
\end{equation}
%
%
\item a right-moving hole that is reflected into a left-moving hole [normal reflection] and a right-moving electron [Andreev reflection] and transmitted transmitted into the superconductor as a linear combination of a right-moving quasi-electron and right-moving quasi hole:
%
%
\begin{equation}\label{P2}
\left.\psi_h^\rightarrow + r'_\text{N}\psi_h^\leftarrow+r'_\text{A}\psi_e^\leftarrow\right|_{x=0}=\left.a'\psi_\text{QE}^\rightarrow+b'\psi_\text{QH}^\rightarrow\right|_{x=0}\,.
\end{equation}
%
%
\end{enumerate}
%
%
We can solve the two processes separately, so to obtain the reflection amplitude $r_\text{N},r'_\text{N},r_\text{A},r_\text{A}'$ that have been show in Eqs.~\eqref{results} --- the corresponding scattering matrix is defined in Eq.~\eqref{sm:NS}.

\section{Solution of the heat balance equation}\label{rate:eq}

In the text we calculate the effective electron temperature solving the following equation
\begin{align}\label{cooling_temperature1_app}
\mathcal{J}(T_e,T_\text{ph},V)+\Sigma \mathcal{V}(T_e^5-T_\text{ph}^5)=0
\end{align}
%
%
where 
%
%
\begin{widetext}
\begin{align}\label{cooling_temperature2_app}
\mathcal{J}(T_e,T_\text{ph},V)=& \frac{1}{\pi \hbar} \int_{-\infty}^\infty d\epsilon \int_0^{\pi/2} d\alpha \cos\alpha \,\,\mathcal{N}(\epsilon)\left\{ \left[\epsilon(1-|r_\text{A}(\epsilon,\alpha)|^2-|r_\text{N}(\epsilon,\alpha)|^2)\right]\right. - \nonumber \\
&\left. eV\left[\left(1+|r_\text{A}(\epsilon,\alpha)|^2-|r_\text{N}(\epsilon,\alpha)|^2\right)\right]\right\} \left[f_\text{N}(\epsilon-eV,T_e)-f_\text{S}(\epsilon,T_\text{ph})\right]\,.
\end{align}
%
%
In our numerical routine we transform Eq.~\ref{cooling_temperature1_app} into:
%
%
\begin{equation}
\label{cooling_temperature2_app}
\frac{\Delta^2 G_0(0)} {e^2 T_\text{c}^5} J\left(\frac{T_e}{T_\text{c}},\frac{T_\text{ph}}{T_\text{c}},\frac{V}{\Delta}\right)= \Sigma \mathcal{V}\left(\left(\frac{T_\text{ph}}{T_\text{c}}\right)^5-\left(\frac{T_e}{T_\text{c}}\right)^5\right)
\end{equation}
%
%
where $J(T_e,T_\text{ph},V)$ is the dimensionless integral
%
%
\begin{align}\label{cooling_temperature3_app}
J(T_e,T_\text{ph},V)=&   \int_{-\infty}^\infty d x \int_0^{\pi/2}\!\!\!\! d\alpha \cos\alpha  \left\{ \left[\frac{x (\mu/\Delta+x) }{\mu/\Delta}(1-|r_\text{A}(x,\alpha)|^2-|r_\text{N}(x,\alpha)|^2)\right]\right. - \nonumber \\
&\hspace{-2cm}\left. eV/\Delta\left[\left(1+|r_\text{A}(x,\alpha)|^2-|r_\text{N}(x,\alpha)|^2\right)\right]\right\} \left[f_\text{N}\left(x-\frac{eV}{\Delta},\frac{T_e}{T_\text{c}}\right)-f_\text{S}\left(x,\frac{T_\text{ph}}{T_\text{c}}\right)\right]\,.
\end{align}
\end{widetext}
%
%
hence the ratio between the two prefactors in the r.h.s. and l.h.s. of Eq. \eqref{cooling_temperature2_app} is our dimensionless control parameter  $\frac{\Sigma \mathcal{V}T_c^5 e^2}{ \Delta^2 G_0(0)} \sim 1 $ in the case of 100 conducting channels and $\frac{\Sigma \mathcal{V}T_c^5 e^2}{ \Delta^2 G_0(0)} \sim 5 $ in the case of 20 conducting channels.

\end{document}